\documentclass{emulateapj}

\def\lesssim{\mathrel{\hbox{\rlap{\hbox{\lower4pt\hbox{$\sim$}}}\hbox{$<$}}}}
\def\gtsim{\mathrel{\hbox{\rlap{\hbox{\lower4pt\hbox{$\sim$}}}\hbox{$>$}}}}

\def\nh3{\mbox{${\rm NH_3}$}}

\begin{document}

\title{Photodesorption of CO ice}

\author{Karin I. \"Oberg$^{1\ast}$, Guido W. Fuchs$^{1}$, Zainab Awad$^{1}$,
Helen J. Fraser$^{2}$, Stephan Schlemmer$^{3}$, Ewine F. van Dishoeck$^{4}$ and
Harold Linnartz$^{1}$
}

\altaffiltext{$^\star$}{To whom correspondence should be addressed; E-mail:  oberg@strw.leidenuniv.nl.}
\altaffiltext{1}{Sackler Laboratory for Astrophysics, Leiden Observatory, University of Leiden, P.O. Box 9513, NL 2300 RA Leiden, The Netherlands.}
\altaffiltext{2}{Department of Physics, Scottish Universities Physics Alliance (SUPA), University of Strathclyde, John Anderson Building, 107 Rottenrow East, Glasgow G4 ONG, Scotland.}
\altaffiltext{3}{Physikalisches Institut, Universit\"at zu K\"oln, Z\"ulpicher Str. 77, 50937 Cologne, Germany}
\altaffiltext{4}{Leiden Observatory, University of Leiden, P.O. Box 9513, NL 2300 RA Leiden, The Netherlands.}


\begin{abstract}
At the high densities and low temperatures found in star forming regions, all molecules other than
H$_2$ should stick on dust grains on timescales shorter than the
cloud lifetimes. Yet these clouds are detected in the millimeter lines of
gaseous CO.  At these temperatures, thermal desorption is negligible and hence a non-thermal desorption mechanism is necessary to maintain molecules in the gas phase. Here, the first laboratory study  of
the photodesorption of pure CO ice under ultra high vacuum is presented, which gives a
desorption rate of $3 \times 10^{-3}$ CO molecules per UV (7--10.5 eV) photon at 15
K. This rate is factors of 10$^2$-10$^5$ larger than previously estimated and is comparable to estimates of other non-thermal desorption rates. The experiments constrains the mechanism to a single photon desorption process of ice surface molecules. The measured efficiency of this process shows that the role of CO photodesorption in preventing total removal of molecules in the gas has been underestimated.
\end{abstract}


\section{Introduction}

In the cold and dense interstellar regions in which stars are formed, CO and other molecules collide with and stick to cold (sub)micron-sized silicate particles, resulting in icy mantles \citep{Leger85,Boogert04}. Chemical models of these regions show that all molecules except for H$_2$ are removed from the gas phase within $\sim 10^9 / n_{\rm H}$ years, where $n_{\rm H}$ is the total hydrogen number density \citep{Willacy98}. For a typical density of $10^4$ cm$^{−3}$, this time scale is much shorter than the estimated age of such regions and hence molecules like CO should be completely frozen out in these clouds. Yet, these clouds are detected in the millimeter lines of gaseous CO \citep{Bergin01, Bergin02}. Similarly cold CO gas has been detected in the midplanes of protoplanetary disks \citep{Dartois03}. A recent study of several disks \citep{Pietu07} even finds that the bulk of the gaseous CO is at temperatures lower than 17 K, below the condensation temperature onto grains. Thus some desorption mechanism is needed to keep part of the CO and other molecules in the gas phase. Clarifying this desorption mechanism is important in understanding the physical and chemical evolution of interstellar clouds. Because the sticking probability of even volatile species like CO has been shown to be unity \citep{Bisschop06}, it is the desorption mechanism and rate that controls the allocation of molecules between gas and solid phase. This allocation of molecules affects the gas phase and surface reactions as well as the dust properties. 

The case of CO is of particular importance, as it is the most common molecule after H$_2$ and the prime tracer of molecular gas. It is also a key constituent in the formation of more complex and pre-biotic species \citep{Tielens97}, and its partitioning between the grain and gas phase therefore has a large impact on the possible chemical pathways \citep{Vandishoeck06b}. In dense clouds without embedded energy sources, the grain temperature is low enough, around 10 K, that thermal desorption is negligible and hence desorption must occur through photon or cosmic ray induced processes. External UV photons from the interstellar radiation field can penetrate into the outer regions of dense clouds and cosmic rays are always present, even in the most shielded regions.

Photodesorption has been proposed as an important desorption pathway of ices in protoplanetary disks and other astrophysical regions with dense clumps of material and excess UV photons \citep{Willacy00,Dominik05}. The lack of experimentally determined photodesorption rates for most astrophysically relevant molecules and conditions has prevented progress in this area, however, and theoretical estimates range by orders of magnitude, with desorption rates from 10$^{-5}$ to 10$^{-8}$ CO molecules UV-photon$^{-1}$ \citep{Draine79, Hartquist90}. Due to this low estimated rate, CO photodesorption has generally been regarded as an insignificant process in astrophysical environments. 

As the present study shows, this assumption is not correct, and the actual desorption rate is at least two orders of magnitude larger than the previous high estimate. Here, we present the results of an experimental study under astrophysically relevant conditions of the photodesorption rate of CO ice and of the mechanism involved. 

\section{Experiments}\label{sec:experiments}

The experimental set-up has been described in detail elsewhere \citep{Fuchs06}. In these experiments, thin ices of 2 to 350 monolayers (ML) are grown at 15~K on a gold substrate under ultra-high vacuum conditions ($P<10^{-10}$ mbar). The ice films are subsequently irradiated at normal incidence with UV light from a broadband hydrogen microwave discharge lamp, which peaks around 125 nm and covers
120--170 nm (7--10.5 eV) \citep{Munozcaro03}. The lamp has a UV photon flux, measured with a NIST calibrated silicon
diode, of $(6\pm2)\times 10^{13}$ photons s$^{-1}$ cm$^{-2}$ at the
substrate surface in its standard setting. The emission resembles the spectral distribution of
the UV interstellar radiation field that impinges externally on all
clouds as well as that of the UV radiation produced locally inside
clouds by the decay of electronic states of H$_2$, excited by energetic
electrons resulting from cosmic-ray induced ionization of hydrogen
\citep{Sternberg87}.
The setup allows simultaneous
detection of molecules in the gas phase by quadrupole mass
spectrometry (QMS) and in the ice by reflection absorption infrared
spectroscopy (RAIRS) using a Fourier transform infrared spectrometer.

Once an ice is grown, it remains stable until it is UV
irradiated. The layer thickness of the ice is monitored by recording RAIR spectra
(Fig. 1). The intensity of the CO RAIRS profile is linearly
correlated with the layer thickness of the CO ice up to $\sim$20
monolayers (ML). One monolayer is generally taken to consist of $\sim$10$^{15}$
molecules cm$^{-2}$ and the rate of the CO photodesorption is
subsequently derived from the intensity loss in the RAIR spectra as
function of time (Fig. 2). From this loss of ice molecules
and the known photon flux hitting the surface it is possible to calculate the
desorption rate as the number of molecules desorbed per incident
photon. Re-condensation will play a negligible role given the small surface
area of the sample and the resulting unterestimate of the actual
photodesorption will be substantially lower than other sources of
inaccuracy. Above 20 ML the photodesorption rate can no
longer be calculated from the RAIRS profile as the integrated
absorbance of the peak is no longer linearly dependent on the number
of molecules. Instead a relative photodesorption rate can be
determined by mass spectrometry of the desorbed gas phase
molecules. This rate is converted to an absolute photodesorption rate
by comparison with thin layer experiments where both QMS and RAIRS
data are available.  Simultaneous mass spectrometry of gas phase
constituents shows that only CO molecules are desorbed. Furthermore
the RAIRS results show that no other molecules are formed during the
UV irradiation (i.e. less than 0.2\% of the CO ice is converted to CO$_2$ after 8 hours of irradiation of 8 ML CO ice). This result
is of importance as in traditional vacuum experiments with
substantially thicker and less pure ices, reaction products have been
identified upon UV photoprocessing \citep{Loeffler05} and this may
affect the photodesorption efficiency. 

In these experiments the thickness of the ice, which is needed to determine the desorption rate, was calculated  from the observed difference in desorption from multilayer coverages (constant rate) and monolayer coverages (decreasing rate). From the RAIR spectra at this turning point, the integrated absorbance of 1 ML is estimated to within 20\%. The original thickness of the ice can then be calculated from the integrated absorbance of the RAIRS feature before onset of desorption. This technique is based on the assumption that the ice is quite flat, which is confirmed by the results of the experiments.

\begin{figure}[h]
\resizebox{\hsize}{!}{\includegraphics{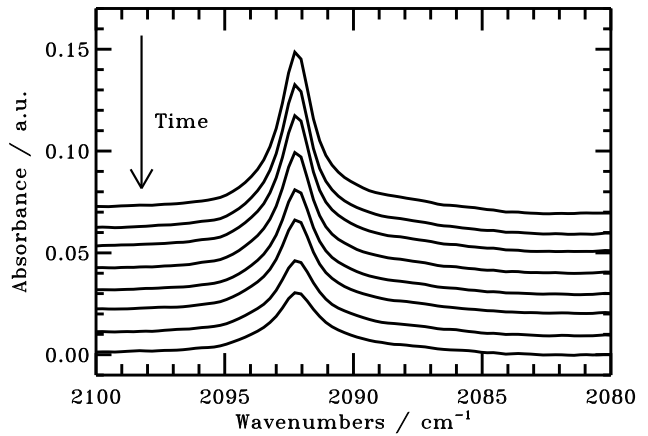}}
\caption{RAIR spectra of the C$^{18}$O $\nu=1-0$ vibrational band at 2040 cm$^{-1}$ (2140 cm$^{-1}$ for normal CO) acquired before irradiation of a 8 ML of C$^{18}$O
ice and then after every hour of irradiation during 8 hours. The drop
in integrated absorbance of the C$^{18}$O ice band is linear with UV irradiation time. In most
of our experiments we used the C$^{18}$O isotopologue instead of
C$^{16}$O to rule out any unwanted contributions from outgassing of the vacuum
chamber. Control experiments with C$^{16}$O resulted in the same
photodesorption rate within our experimental uncertainty.}
\label{fig1}
\end{figure}

\begin{figure}[h]
\resizebox{\hsize}{!}{\includegraphics{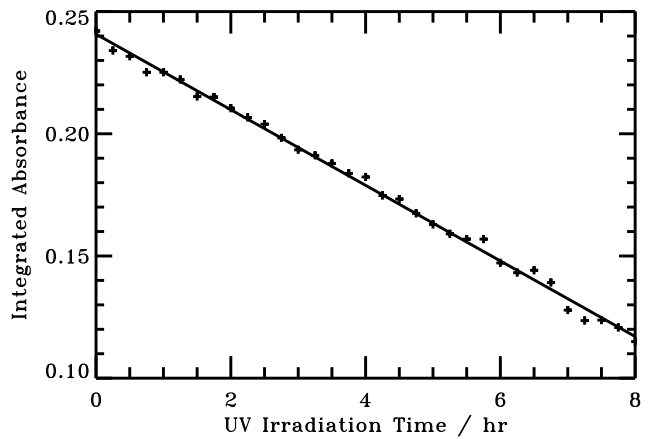}}
\caption{The integrated absorbance of the CO RAIRS band acquired
before irradiation of a 8 ML C$^{18}$O ice and then after every 15
minutes of UV irradiation during 8 hours. The photodesorption rate is calculated from the slope of the fitted line. The fitted line gives a photodesorption rate in loss of integrated absorbance in integrated absorbance units (I.A.U.) per hour (here 0.015 I.A.U. hr$^{-1}$). The amount of integrated absorbance per monolayer can be derived from the integrated absorbance of 0.24 I.A.U. of 8 ML at time 0. Using the known UV flux and CO coverage, the loss of integrated absorbance per hour is converted to a photodesorption rate in CO molecules per UV (7-10.5 eV) photon: ${R_{\rm pd}=\left({0.015  \: \rm I.A.U.}/{\rm hr}\right) \times \left({\rm hr}/{3600 \rm \: s}\right) \times \left( {8 \: \rm ML}/{0.24 \: \rm I.A.U.}\right)}$ ${\times \left({10^{15} \: \rm molecules \: cm^{-2}}/{1 \rm \: ML}\right)\times \left(1/{\rm 6{\times}10^{13} \: photons \: s^{-1} cm^{-2}}\right)}$ ${= \: 0.003 \rm \: molecules \: photon^{-1}}$}
\label{fig2}
\end{figure}

\section{Results}

The evaluation of the experiments results in a constant rate of (3$\pm$1) $\times 10^{-3}$ CO
molecules photon$^{-1}$, averaged over the wavelength range of the lamp. This
rate is fully reproducible from repeated experiments of 8 ML coverage
and has a standard deviation of $\sim$15\% (Fig. 3). The uncertainty in the
absolute value is somewhat larger, up to 30\%, dominated by the
uncertainty in the UV photon flux and coverage.

The thickness of the CO ice has been varied between 2 and 350 ML. We
find that the photodesorption rate of CO is independent of the ice
thickness (Fig. 3). This suggests that only the upper layers are involved in the photodesorption event. It is also consistent with a surface that is quite smooth, since at 2 ML the entire surface must still be covered to achieve the same photodesorption rate as for 350 ML. To confirm that the desorbed molecules only originate from the top layers we
performed experiments with two layers of ices comprising different CO
isotopologues. When 2 ML of C$^{18}$O is deposited on top of 8 ML of
C$^{16}$O the desorption rate from the bottom layer drops with less than 20\%. In contrast depositing 4 ML of C$^{18}$O on top of 8 ML of
C$^{16}$O ice reduces the C$^{16}$O desorption rate with more than 80\%. This
confirms that mainly the top few layers of the CO ice are directly
involved. Furthermore, the CO
photodesorption rate is directly proportional to the photon
flux within the flux range covered here ($(4-8)\times 10^{13}$ photons s$^{-1}$ cm$^{-2}$).

In contrast to previous findings on H$_2$O photodesorption \citep{Westley95}, we find that the CO photodesorption rate is
independent of the total photon dose as well as the irradiation time,
as long as 1 ML is left on the surface. The
mass signals show an onset of the photodesorption within the time constant of our QMS system (a few seconds) when the UV source is turned on. 

In addition to the experiments on CO ices, a thin layer of N$_2$ ice (8 ML) was irradiated under the same conditions as the CO ices with the aim to compare the two photodesorption rates. It is found that N$_2$ has no
detectable photodesorption in the present set-up, which puts an upper bound to
the photodesorption rate of pure N$_2$ ice of 2 x 10$^{-4}$ molecules UV-photon$^{-1}$.

\begin{figure}[h]
\resizebox{\hsize}{!}{\includegraphics{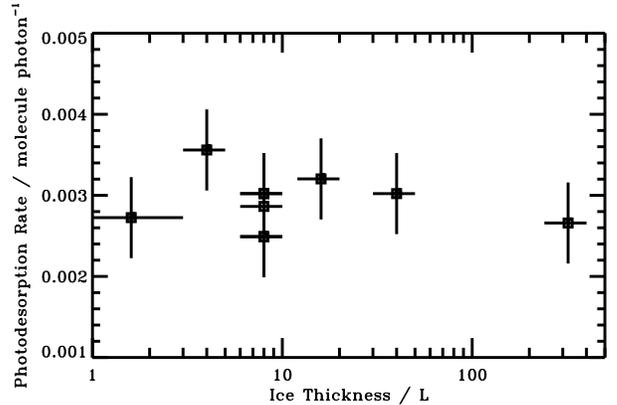}}
\caption{The desorption rate of CO at different layer
thicknesses. From repeated experiments around 8 ML the standard
deviation in the photodesorption rate was determined, indicated by
the size of the error bars. Within the experimental
uncertainty derived from this spread, we conclude that the CO photodesorption rate is independent of the
thickness of the CO ice.}
\label{fig3}
\end{figure}
\newpage 

\section{Discussion}\label{sec:discussion}

\subsection{Photodesorption Mechanism}

The above experiments can be used to constrain the CO photodesorption mechanism. Its insensitivity to layer thickness (demonstrated in
Fig. 3) indicates that only molecules from the top layers of
the ice contribute to the photodesorption flux. In addition, it suggests
that the CO photodesorption mechanism at these UV wavelengths does
not involve the substrate. The linear dependence of the photodesorption rate on the UV intensity is not consistent with that the desorption is due to sublimation caused by heating of the ice as a whole. Together these results suggest that CO photodesorption occurs through a single photon-process, which is further supported by the immediate on-set of the desorption once the UV lamp is turned on. The opposite conclusion has previously been drawn for H$_2$O \citep{Westley95}.

The final support for a single-photon process is given by the different desorption rates for CO and N$_2$ ice. The two molecules have similar inter-molecular binding energies and ice structures \citep{Fuchs06}, which suggests that any difference in photodesorption rate must involve the internal structure of the molecules. A relevant difference between the
two species is that CO has an electric dipole allowed transition in
the vacuum ultraviolet (7--10 eV) \citep{Mason06}, exactly where the
hydrogen lamp simulates the interstellar radiation field, while N$_2$
does not. This transition corresponds to the
solid state equivalent of the A$^1{\Pi}$-$X^{1}{\Sigma}^+$ electronic excitation of gaseous CO and the most plausible photodesorption mechanism hence involves
this transition. After UV absorption, the excited molecule relaxes via
a radiationless transition
into vibrationally excited states of the electronic ground state which
subsequently transfer part of this intramolecular energy to the weak
intermolecular bonds with neighboring CO molecules, resulting in a
desorption event. This desorption event may consist of more than the originally excited molecule desorbing, but experiments with mixed CO/N$_2$ ices are necessary to constrain this part of the mechanism in more detail.

\subsection{Astrophysical Implications}

The single photon mechanism of CO photodesorption means that the rate derived from these experiments can be easily applied to astrophysical environments without concerns about ice thicknesses and irradiation field strengths. It is illustrative to compare the photodesorption rate with other possible desorption routes in dark clouds. 
Since thermal desorption is negligible, species can only be kept in
the gas phase through ice desorption induced by UV photons and cosmic
rays. While cosmic rays have been proposed to directly desorb ices
\citep{Leger85}, the efficiency of heating an entire grain to the
required desorption temperature rapidly drops with grain size so that
usually only spot heating at an estimated rate of 70 molecules
cm$^{-2}$ s$^{-1}$ is considered as a viable mechanism. The cosmic
rays also produce UV photons so
that the total photodesorption rate depends on both the external interstellar radiation field and the UV field produced inside the cloud by the cosmic rays. For a typical galactic cosmic ray
flux, the resulting UV photon flux is of the order of $10^4$
photons cm$^{-2}$ s$^{-1}$ with a factor of 3 uncertainty
\citep{Shen04}. 

We calculated the desorption rate due to photodesorption in a dark cloud and compared this with the desorption due to spot-heating from \citet{Shen04}. Equations \ref{eqn1} and \ref{eqn2} describe the photodesorption rates of CO molecules from grain surfaces in molecules cm$^{-2}$ s$^{-1}$ due to external and cosmic ray induced UV photons, respectively, where $I_{\rm ISRF-FUV}=1\times 10^8 \:\rm
photons \: cm^{-2} \:s^{-1}$ is the strength of the external irradiation field with energies 6-13.6 eV, $I_{\rm CR-FUV}$ the strength of the UV field due to cosmic rays, $\gamma$ is a measure of UV extinction relative to visual extinction, which is $\sim$2 for small interstellar grains \citep{Roberge91}, and $Y_{\rm pd}$ is the experimentally determined photodesorption rate. 

 \begin{equation}
R_{\rm UV-PD}{=}I_{\rm ISRF-FUV}e^{-\gamma A_{\rm V}}Y_{\rm pd}
 \label{eqn1}
 \end{equation}

 \begin{equation}
R_{\rm CR-PD}{=}I_{\rm CR-FUV}Y_{\rm pd}
 \label{eqn2}
 \end{equation}

\begin{figure}[h]
\resizebox{\hsize}{!}{\includegraphics{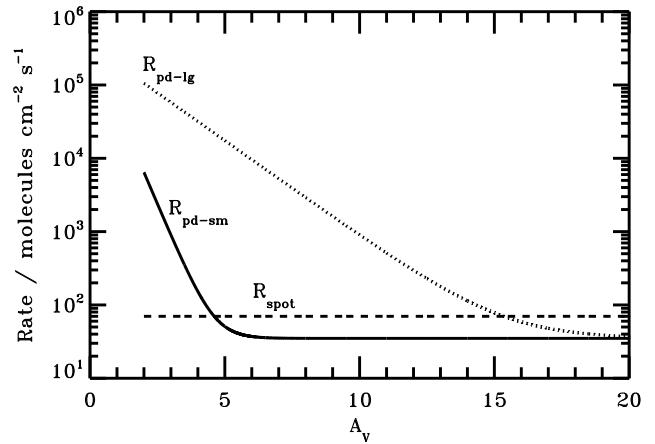}}
\caption{The photodesorption rate of CO for small (0.1 $\mu$m, R$_{\rm pd-sm}$, full line) and large (a few $\mu$m, R$_{\rm pd-lg}$, dotted line) grains compared to desorption due to spot heating by cosmic rays (R$_{\rm spot}$, dashed line). For small grains, applicable to dark clouds, the spot heating and photodesorption rates are comparable. When grains have grown to a few $\mu$m size the photodesorption dominates up to $A_V=15$.}
\label{fig4}
\end{figure}

Applying this model shows that photodesorption dominates at the edge of the cloud and becomes comparable (within the uncertainties of a few) to spot heating in the interior of a cloud, i.e. beyond a depth corresponding to an extinction of 3--4 A$_V$ (Fig. \ref{fig4}). In the interior of the cloud the photodesorption rate due to cosmic rays is $\sim$30 molecules cm$^{-2}$ s$^{-1}$, which is equivalent to $\sim10^{-8}$ molecules grain$^{-1}$ s$^{-1}$ for grains with a 0.1 $\mu$m radius. A rate of this magnitude may on its own explain the gas phase CO seen in dark clouds \citep{Bergin06}. In comparison with the other plausible non-thermal desorption mechanisms, photodesorption has the advantage that the rate can now be determined experimentally and unambiguously included in astrophysical models. 

One particularly interesting application of our derived photodesorption
rate is to the case of CO in protoplanetary disks, where large abundances of cold CO-gas is observed. \citet{Aikawa06} argue that the cold CO-gas can be explained by vertical mixing and \citet{Semenov06} by a combination of radial and vertical mixing of disk material. With the high photodesorption rate of CO derived here, non-thermal desorption of CO may suffice to explain these observations. An important characteristic of disks compared to dense clouds is dust coagulation, which reduces the absorption of the external UV field since larger grains absorb less efficiently at short wavelengths. In Eq. \ref{eqn1} this corresponds to $\gamma \leq 0.6$ \citep{Vandishoeck06}, assuming the grains have
grown to at least $\mu$m size as indicated by infrared observations
of the silicate feature from the surfaces layers of the inner disk
\citep[e.g.][]{Bouwman01} and by millimeter observations of the outer
disk showing growth up to mm size \citep[e.g.][]{Rodmann06}. The photodesorption rate due to external photons then dominates over other non-thermal desorption rates up to A$_V=15$, but detailed modeling is necessary to determine whether this rate is high enough to offer an alternative to the turbulence theory in explaining the observed CO gas. It is clear, however, that photodesorption can no longer be ignored in astrophysical models and may explain a large part of the gas observed in cold and dense regions.

\begin{acknowledgements}
We thank F.A. van Broekhuizen and W.A. Schutte for initial work on our photodesorption instrument. Funding was provided
by NOVA, the Netherlands Research School for Astronomy, a grant from the European Early Stage Training Network ('EARA' MEST-CT-2004-504604) and a NWO Spinoza grant.
 \end{acknowledgements}


\bibliographystyle{apj}

\begin{thebibliography}{25}
\expandafter\ifx\csname natexlab\endcsname\relax\def\natexlab#1{#1}\fi

\bibitem[{{Aikawa} \& {Nomura}(2006)}]{Aikawa06}
{Aikawa}, Y. \& {Nomura}, H. 2006, ApJ, 642, 1152

\bibitem[{{Bergin} {et~al.}(2002){Bergin}, {Alves}, {Huard}, \&
  {Lada}}]{Bergin02}
{Bergin}, E.~A., {Alves}, J., {Huard}, T., \& {Lada}, C.~J. 2002, ApJL, 570,
  L101

\bibitem[{{Bergin} {et~al.}(2001){Bergin}, {Ciardi}, {Lada}, {Alves}, \&
  {Lada}}]{Bergin01}
{Bergin}, E.~A., {Ciardi}, D.~R., {Lada}, C.~J., {Alves}, J., \& {Lada}, E.~A.
  2001, ApJ, 557, 209

\bibitem[{{Bergin} {et~al.}(2006){Bergin}, {Maret}, {van der Tak}, {Alves},
  {Carmody}, \& {Lada}}]{Bergin06}
{Bergin}, E.~A., {Maret}, S., {van der Tak}, F.~F.~S., {Alves}, J., {Carmody},
  S.~M., \& {Lada}, C.~J. 2006, ApJ, 645, 369

\bibitem[{{Bisschop} {et~al.}(2006){Bisschop}, {Fraser}, {{\"O}berg}, {van
  Dishoeck}, \& {Schlemmer}}]{Bisschop06}
{Bisschop}, S.~E., {Fraser}, H.~J., {{\"O}berg}, K.~I., {van Dishoeck}, E.~F.,
  \& {Schlemmer}, S. 2006, A{\&}A, 449, 1297

\bibitem[{{Boogert} \& {Ehrenfreund}(2004)}]{Boogert04}
{Boogert}, A.~C.~A. \& {Ehrenfreund}, P. 2004, in ASP Conf. Ser. 309:
  Astrophysics of Dust, ed. A.~N. {Witt}, G.~C. {Clayton}, \& B.~T. {Draine},
  547--572

\bibitem[{{Bouwman} {et~al.}(2001){Bouwman}, {Meeus}, {de Koter}, {Hony},
  {Dominik}, \& {Waters}}]{Bouwman01}
{Bouwman}, J., {Meeus}, G., {de Koter}, A., {Hony}, S., {Dominik}, C., \&
  {Waters}, L.~B.~F.~M. 2001, \aap, 375, 950

\bibitem[{{Dartois} {et~al.}(2003){Dartois}, {Dutrey}, \&
  {Guilloteau}}]{Dartois03}
{Dartois}, E., {Dutrey}, A., \& {Guilloteau}, S. 2003, A{\&}A, 399, 773

\bibitem[{{Dominik} {et~al.}(2005){Dominik}, {Ceccarelli}, {Hollenbach}, \&
  {Kaufman}}]{Dominik05}
{Dominik}, C., {Ceccarelli}, C., {Hollenbach}, D., \& {Kaufman}, M. 2005, ApJL,
  635, L85

\bibitem[{{Draine} \& {Salpeter}(1979)}]{Draine79}
{Draine}, B.~T. \& {Salpeter}, E.~E. 1979, ApJ, 231, 438

\bibitem[{Fuchs {et~al.}(2006)Fuchs, Acharyya, Bisschop, \"Oberg, van
  Broekhuizen, Fraser, Schlemmer, van Dishoeck, \& Linnartz}]{Fuchs06}
Fuchs, G.~W., Acharyya, K., Bisschop, S.~E., \"Oberg, K. .~I., van Broekhuizen,
  F.~A., Fraser, H.~J., Schlemmer, S., van Dishoeck, W.~F., \& Linnartz, H.
  2006, Faraday Discussions, 133, 331

\bibitem[{{Hartquist} \& {Williams}(1990)}]{Hartquist90}
{Hartquist}, T.~W. \& {Williams}, D.~A. 1990, Mon. Not. R. astr. Soc., 247, 343

\bibitem[{{L{\'e}ger} {et~al.}(1985){L{\'e}ger}, {Jura}, \& {Omont}}]{Leger85}
{L{\'e}ger}, A., {Jura}, M., \& {Omont}, A. 1985, A{\&}A, 144, 147

\bibitem[{Loeffler {et~al.}(2005)Loeffler, Baratta, Palumbo, Strazzulla, \&
  Baragiola}]{Loeffler05}
Loeffler, M.~J., Baratta, G.~A., Palumbo, M.~E., Strazzulla, G., \& Baragiola,
  R.~A. 2005, A{\&}A, 435, 587

\bibitem[{Mason {et~al.}(2006)Mason, Dawes, Holtom, Mukerji, Davis, Sivaraman,
  Kaiser, Hoffmann, \& Shaw}]{Mason06}
Mason, N.~J., Dawes, A., Holtom, P.~D., Mukerji, R.~J., Davis, M.~P.,
  Sivaraman, B., Kaiser, R.~I., Hoffmann, S.~V., \& Shaw, D.~A. 2006, Faraday
  Discussions, 133, 1

\bibitem[{{Mu{\~n}oz Caro} \& {Schutte}(2003)}]{Munozcaro03}
{Mu{\~n}oz Caro}, G.~M. \& {Schutte}, W.~A. 2003, A{\&}A, 412, 121

\bibitem[{{Pi{\'e}tu} {et~al.}(2007){Pi{\'e}tu}, {Dutrey}, \&
  {Guilloteau}}]{Pietu07}
{Pi{\'e}tu}, V., {Dutrey}, A., \& {Guilloteau}, S. 2007, ArXiv Astrophysics
  e-prints

\bibitem[{{Roberge} {et~al.}(1991){Roberge}, {Jones}, {Lepp}, \&
  {Dalgarno}}]{Roberge91}
{Roberge}, W.~G., {Jones}, D., {Lepp}, S., \& {Dalgarno}, A. 1991, \apjs, 77,
  287

\bibitem[{{Rodmann} {et~al.}(2006){Rodmann}, {Henning}, {Chandler}, {Mundy}, \&
  {Wilner}}]{Rodmann06}
{Rodmann}, J., {Henning}, T., {Chandler}, C.~J., {Mundy}, L.~G., \& {Wilner},
  D.~J. 2006, A{\&}A, 446, 211

\bibitem[{{Semenov} {et~al.}(2006){Semenov}, {Wiebe}, \& {Henning}}]{Semenov06}
{Semenov}, D., {Wiebe}, D., \& {Henning}, T. 2006, \apjl, 647, L57

\bibitem[{{Shen} {et~al.}(2004){Shen}, {Greenberg}, {Schutte}, \& {van
  Dishoeck}}]{Shen04}
{Shen}, C.~J., {Greenberg}, J.~M., {Schutte}, W.~A., \& {van Dishoeck}, E.~F.
  2004, A{\&}A, 415, 203

\bibitem[{{Sternberg} {et~al.}(1987){Sternberg}, {Dalgarno}, \&
  {Lepp}}]{Sternberg87}
{Sternberg}, A., {Dalgarno}, A., \& {Lepp}, S. 1987, ApJ, 320, 676

\bibitem[{{Tielens} \& {Charnley}(1997)}]{Tielens97}
{Tielens}, A.~G.~G.~M. \& {Charnley}, S.~B. 1997, Origins of Life and Evolution
  of the Biosphere, 27, 23

\bibitem[{{van Dishoeck}(2006)}]{Vandishoeck06b}
{van Dishoeck}, E.~F. 2006, Proceedings of the National Academy of Science,
  103, 12249

\bibitem[{van Dishoeck {et~al.}(2006)van Dishoeck, Jonkheid, \& van
  Hemert}]{Vandishoeck06}
van Dishoeck, E.~F., Jonkheid, B., \& van Hemert, M.~C. 2006, Faraday
  Discussions, 133, 231

\bibitem[{{Westley} {et~al.}(1995){Westley}, {Baragiola}, {Johnson}, \&
  {Baratta}}]{Westley95}
{Westley}, M.~S., {Baragiola}, R.~A., {Johnson}, R.~E., \& {Baratta}, G.~A.
  1995, Nature, 373, 405

\bibitem[{{Willacy} \& {Langer}(2000)}]{Willacy00}
{Willacy}, K. \& {Langer}, W.~D. 2000, ApJ, 544, 903

\bibitem[{{Willacy} \& {Millar}(1998)}]{Willacy98}
{Willacy}, K. \& {Millar}, T.~J. 1998, \mnras, 298, 562

\end{thebibliography}

\end{document}